\documentclass[twocolumn,showpacs,superscriptaddress,preprintnumbers,amsmath,amssymb,longbibliography,prl]{revtex4-1}
\usepackage{graphicx}% Include figure files
\usepackage{dcolumn}% Align table columns on decimal point
\usepackage{bm}% bold math
\usepackage{amsmath}
\usepackage{color}
\usepackage{wrapfig}
\begin{document}

\title{Granular temperature controls local rheology of vibrated granular flows}

\author{Mitchell G. Irmer}
\affiliation{Department of Physics, Naval Postgraduate School, Monterey, California 93943, USA}
\author{Emily E. Brodsky}
\affiliation{Department of Earth and Planetary Sciences, University of California Santa Cruz, Santa Cruz, California 95064, USA}
\author{Abram H. Clark}
\affiliation{Department of Physics, Naval Postgraduate School, Monterey, California 93943, USA}

\begin{abstract}

We use numerical simulations to demonstrate a local rheology for sheared, vibrated granular flows. We consider a granular assembly that is subjected to simple shear and harmonic vibration at the boundary. This configuration allows us to isolate the effects of vibration, as parameterized by granular temperature. We find that friction is reduced due to local velocity fluctuations of grains. All data obey a local rheology that relates the material friction coefficient, the granular temperature, and the dimensionless shear rate. We also observe that reduction in material friction due to granular temperature is associated with reduction in fabric anisotropy. We demonstrate that the temperature can be modeled by a heat equation with dissipation with appropriate boundary conditions, which provides complete closure of the system and allows a fully local continuum description of sheared, vibrated granular flows. This success suggests local rheology based on temperature, as suggested previously, combined with the new, empirical heat diffusion equation may provide a general strategy for dense granular flows. 
\end{abstract}

\date{\today}

\maketitle	

Sheared granular materials display friction-like behavior~\cite{bingham1917investigation,drucker1952soil}, but the friction coefficient friction $\mu$ depends on flow speed and boundary conditions in a way that cannot be easily predicted. In limited cases, an inertial rheology $\mu(I)$ can be calibrated to predict flows in complex geometries~\cite{jop2006constitutive} where $I$ is the local inertial number $\dot{\gamma}d\sqrt{\rho/p}$~\cite{daCruz2005}. Here, $\dot{\gamma}$ is the shear rate, $d$ is the grain diameter, $\rho$ is the mass density of each grain, and $p$ is the confining pressure. However, inertial rheology fails in systems with stress gradients~\cite{gdr2004dense, kovalPRE2009, wandersman2014nonlocal} or when vibrations are present, either from external sources \cite{melosh1979acoustic, melosh1996dynamical, Dijksman2011, taslagyan2015effect, ClarkPRL2023} or generated by the flow itself~\cite{vanderElst2012auto, taylor2017granular, degiuli2017friction, taylor2020reversible}. In such cases, $\mu$ is often reduced below the value predicted by a $\mu(I)$ rheology, given the observed $I$. 

There is growing evidence that granular temperature $T = (\delta v)^2$, where $\delta v$ is a measure of local velocity fluctuations, may be a crucial to understanding deviations from $\mu(I)$ rheology. Experiments have shown that measured temperature is correlated with reduced $\mu$~\cite{vanderElst2012auto,taylor2020reversible}. Additionally, models based on nonlocal fluidity can predict deviations from $\mu(I)$ due to stress gradients~\cite{kamrin2012nonlocal,henann2013predictive,bouzid2013nonlocal,bouzid2015non} and distance to boundaries~\cite{Kamrin2015SoftMatt_hstop}, but these effects are also consistent with a local rheology that includes dimensionless temperature $\Theta = \rho T/p$~\cite{Zhang2017PRL,KimPRL2020}. \citet{Berzi2024PRF} demonstrates that many aspect of fluidity can be derived from granular temperature as described by kinetic theory. Although granular temperature may be more easily interpretable than fluidity, there is currently no governing equation for the propagation of temperature in a dense granular flow and thus the system is currently not closed.  Additionally, there are many open questions about the micro- and meso-structural mechanisms that reduce $\mu$, the answers to which may aid in the development of a successful modeling framework for temperature.

In this Letter, we address these questions by seeking a local rheology for sheared, vibrated granular flows that includes granular temperature. We use discrete element method (DEM) simulations with vibration at the boundary of amplitude $A$ and frequency $f$. We make local, time-averaged measurements of $\mu$, $I$, and $\Theta$ for varied $A$ and $f$ and find that a relation $\mu(I,\Theta)$ emerges. When dimensionless acceleration $\Gamma>1$, the vibrating wall acts as a temperature source, with a boundary condition set by the peak wall velocity. Away from the wall, temperature varies spatially in a way that is consistent with a modified heat equation similar to others developed using granular kinetic theory, with key differences that we discuss. Perhaps most importantly, we find a local rheology similar to the one observed in the context of the nonlocal granular fluidity model~\cite{KimPRL2020}. This suggests that that a single \textit{local} rheology may be sufficient to capture any steady, dense granular flow including vibrations, as we explicitly study, as well as stress gradients and finite size effects. We also show that $\Theta\ll 1$ for the vast majority of our data, meaning the contact stresses are still dominant. In this case, contact and force fabric tensors primarily control $\mu$, which is reduced as temperature destroys fabric anisotropy.

\textit{Methods---}We follow our previous work~\cite{ClarkPRL2023}, which demonstrated that a system-averaged $\mu$ is reduced due to wall vibrations that must satisfy two conditions. First, dimensionless acceleration $\Gamma = A (2\pi f)^2 \rho d/p > 1$ in order to break contacts with particles~\cite{Dijksman2011}. Second, the amplitude $\tilde{A} = A/d$ must be large enough (relative to the pressure) to disrupt the contact networks that are responsible for the frictional response. Here we use similar simulations but take local fmeasurements of $\mu$, $I$, and $\Theta$ for varying $\Gamma$ and $\tilde{A}$.

We use LAMMPS~\cite{LAMMPS} to simulate $N = 1000$ spherical grains of average diameter $d$ via the motion of a top wall with imposed vibrations at the bottom wall. Both top and bottom walls are constructed from rigidly connected grains. The horizontal dimensions are both periodic with length $L = 7d$ or $L = 10d$, corresponding to a height of roughly $20d$ or $10d$, respectively; results from both are typically indistinguishable, and data shown uses $L = 10$, height of $20d$, unless otherwise stated. Intergrain forces are Hookean and characterized by a modulus $E$, normal restitution coefficient $e_n$, and friction coefficient $\mu_g$; see Supplemental Material for further details~\cite{suppmat}. We impose a normal stress $p$ and a horizontal velocity $v$ on the top wall. We use $\tilde{p} \equiv p/E \leq 10^{-3}$, and we verify our results are identical for $\tilde{p} = 10^{-3}$ and $10^{-4}$. On the bottom wall we impose a vertical harmonic displacement with amplitude $A$ and frequency $f$.

\begin{figure}
    \raggedright (a) \\
    \centering
    \includegraphics[trim=35mm 40mm 35mm 30mm, clip,width=0.49\columnwidth]{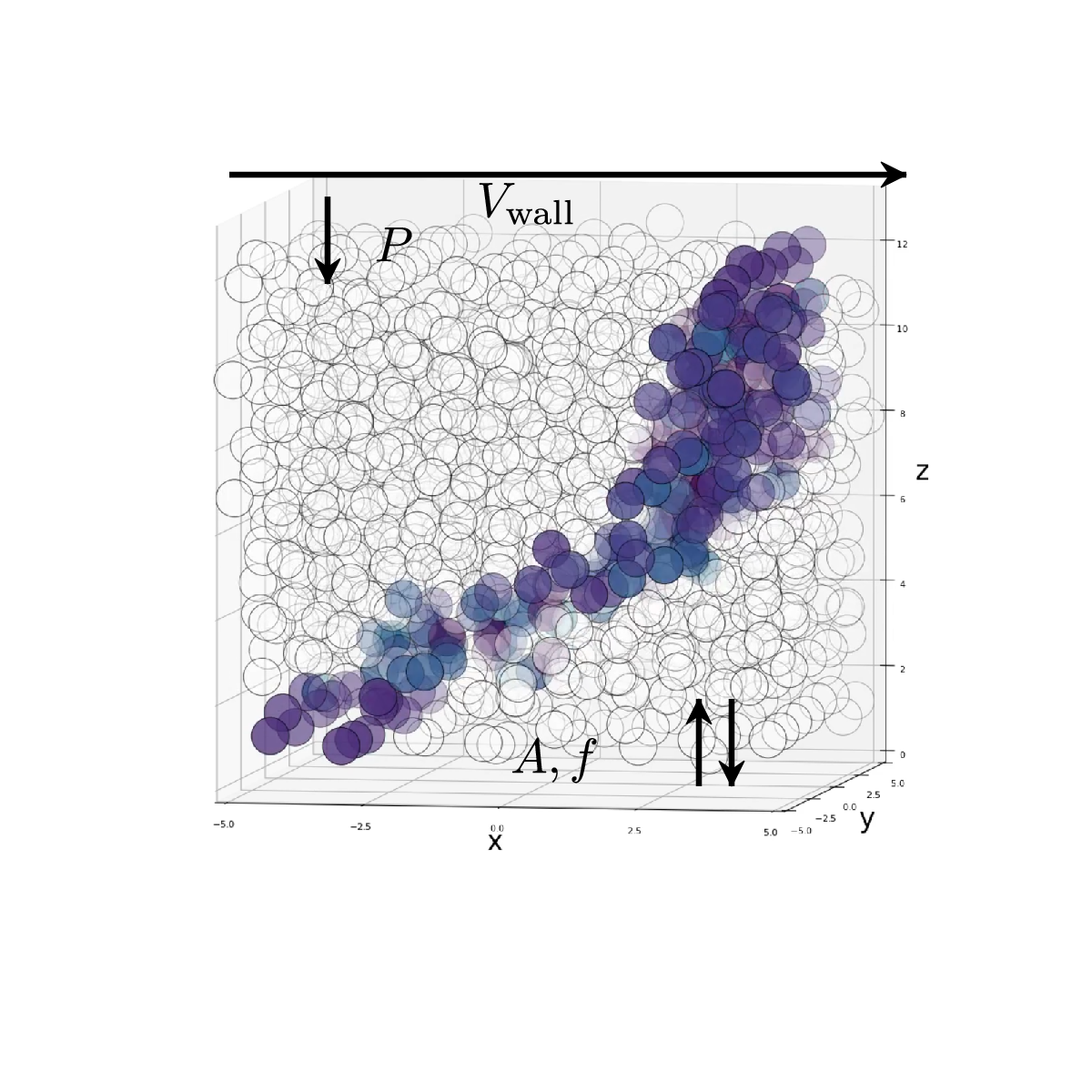}
    \includegraphics[trim=35mm 40mm 35mm 30mm, clip,width=0.49\columnwidth]{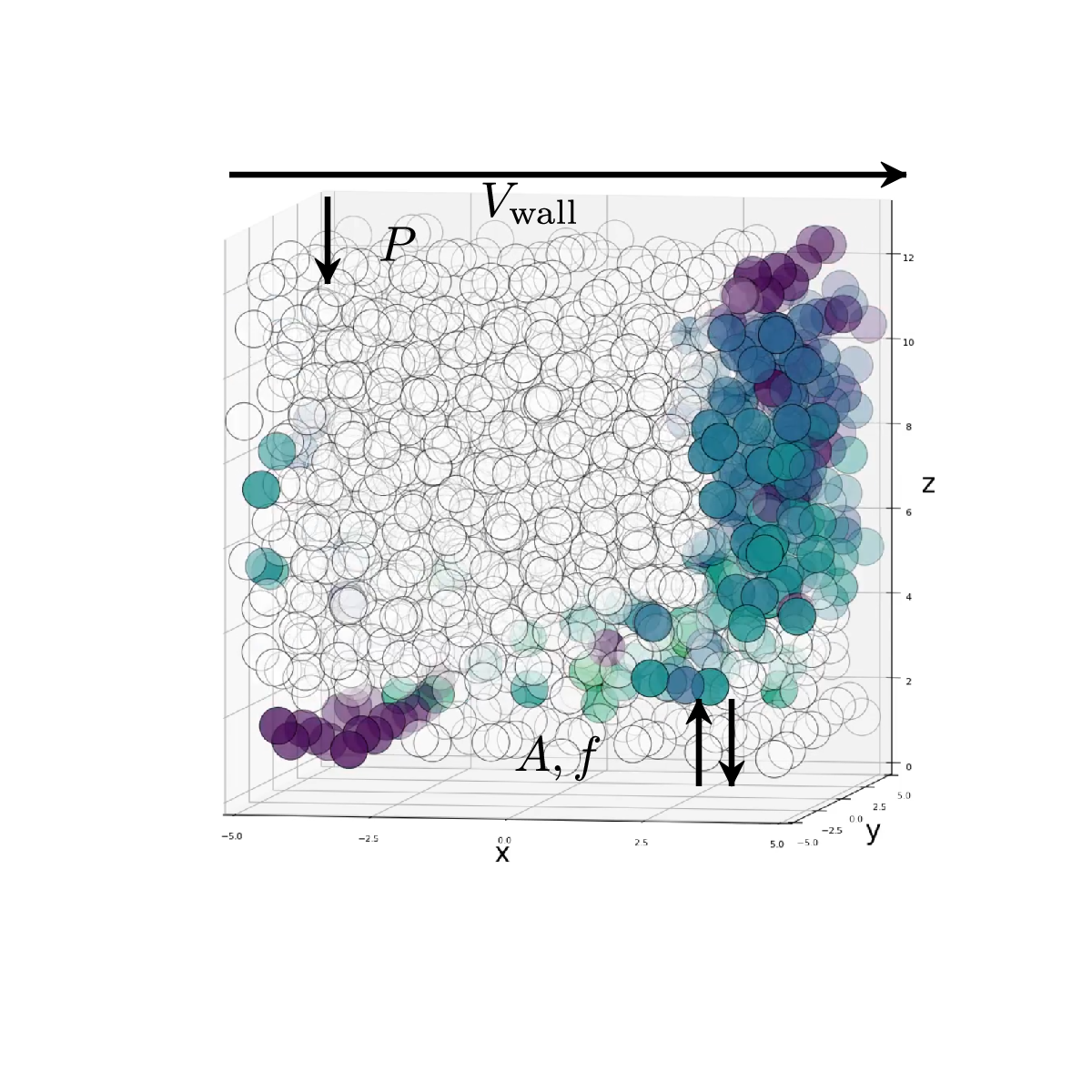}

    \raggedright (b) \\
    \includegraphics[trim=0mm 0mm 0mm 0mm,clip,width=\columnwidth]{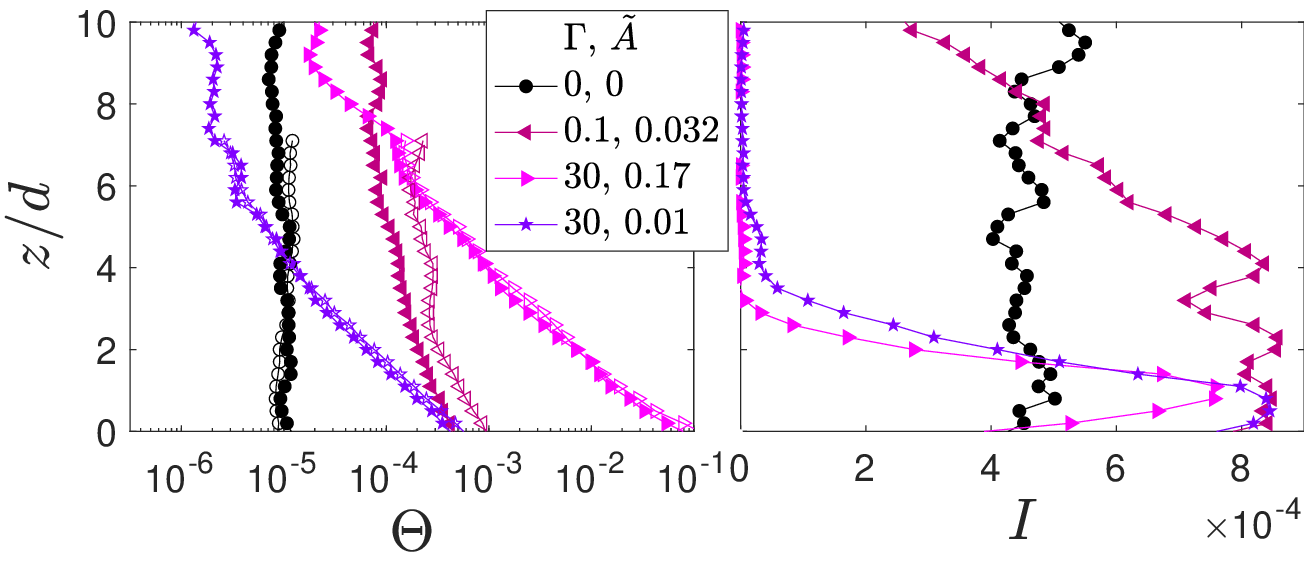} \caption{(a) Snapshot of simulations with $\Gamma = 0.1,\ \tilde{A} = 0.01$ (left) and $\Gamma = 30,\ \tilde{A} = 0.1$ (right), after global strain of $\gamma = 1$. Colored particles all started at $x\approx 0$ and show the average velocity profile; brighter colors correspond to higher granular temperatures.  (b) Representative profiles of $\Theta$ and $I$ for the same top wall velocity with several values of the vibration parameters $\Gamma$ and $\tilde{A}$; height 10$d$ (open symbols) and 20$d$ show good agreement.}
    \label{fig:cartoon}
\end{figure}

We collect data over global strains of $\gamma \approx 7$ after applying global strain of 1 to ensure transients have decayed. For each simulation, we obtain time series of grain positions and velocities; layer-averaged stress tensor elements $\sigma_{\alpha \beta}$, including kinetic and contact stresses; and layer-averaged fabric anisotropy elements $a_c$, $a_n$, and $a_t$, described further below. We use the grain velocities to compute layer-averaged, instantaneous temperature profiles for the full system. The temperature in each layer is $T \approx T_{yy} = \overline{ (v_y-\bar{v}_y)^2}$. These instantaneous profiles are then time-averaged to obtain a profile of temperature as a function of height. Velocity profiles are also computed by averaging by layer and then over time, and $\dot{\gamma}$ is measured as $dv_x/dz$, which then yields $I(z)$. The friction coefficient $\mu \approx \langle \sigma_{xz}/\sigma_{zz} \rangle$ is computed in three layers centered at $z=2d$, $4d$, and $6d$, where angle brackets denote a time average. We verify that our results are all insensitive to the thickness of the layers used for spatial averaging (typically of size $d$).

\textit{Temperature field observations---}We first discuss and physically interpret the temperature fields resulting from varied $\Gamma$ and $\tilde{A}$. For no vibration ($\Gamma = \tilde{A} = 0$) the system is homogeneous, and $I$ and $\Theta$ are both constant as a function of distance from the wall $z/d$ (black curves in Fig.~\ref{fig:cartoon}(b)). In this case, temperature is strictly generated by shear, as particles rearrange and move past each other. For nonzero $\Gamma$ and $\tilde{A}$, the wall also generates $\Theta$ and we observe significant variation in both $I(z)$ and $\Theta(z)$. When $\Gamma > 1$, the vibrating wall breaks contacts with the particles. In this regime, the $\Theta(z)$ profiles decay away from the wall quasi-exponentially until reaching a plateau, and $I(z)$ also varies; this highlights the need for a local description. For $\Gamma < 1$, $\Theta$ increases above the zero-vibration baseline, but there is very little change in $I(z)$, as seen on the left of Fig.~\ref{fig:cartoon}(a) as well as the dark red curves in Fig.~\ref{fig:cartoon}(b). We interpret this as fluctuating particle velocities due to elastic vibrational modes being driven by slow wall vibrations. This type of $\Theta$ increase does not lead to reduction in $\mu$, as shown in Supplemental Material~\cite{suppmat}. For $10^{-2}<\Gamma<1$, some weakening is observed. For the remainder of the paper, we focus on the $\Gamma > 1$ regime, corresponding to when the vibrating wall breaks contacts with the particles.

\begin{figure}
    \centering
    \includegraphics[trim=0mm 0mm 0mm 0mm, clip, width=\columnwidth]{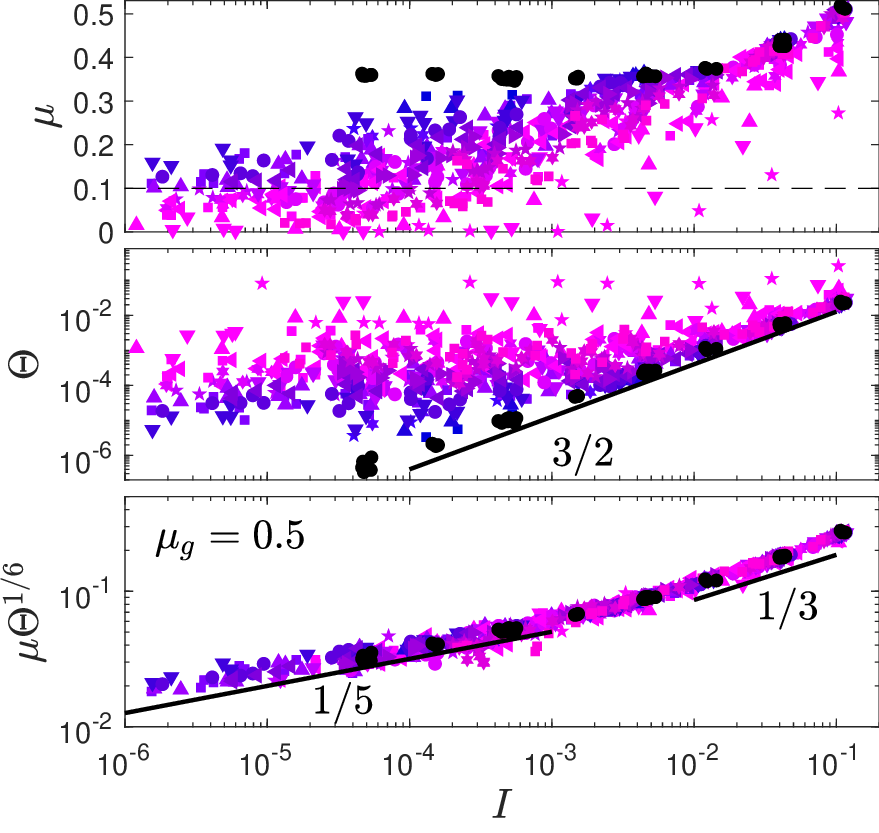}\\
    \centering
    \caption{Local, time-averaged measurements of $\mu$, $\Theta$, and $\mu \Theta^{1/6}$ versus $I$ for many simulations with $3\leq \Gamma \leq 100$ and varied $\tilde{A}$ and wall velocity. Black symbols correspond to no vibration. Symbol color (blue to pink) signifies increasing vibration amplitude. Data for $\mu \Theta^{1/6}$ excludes any data with $\mu<0.1$ below the dashed black line in the top panel.}
    \label{fig:fric_collapse}
\end{figure}

\textit{Local rheology observations---} Figure~\ref{fig:fric_collapse} shows time-averaged, local $\mu(I)$ and $\Theta(I)$ with $\mu_g = 0.5$. The black symbols correspond to no vibration, which reproduces the local rheology $\mu(I)$ and the relationship $\Theta \propto I^{3/2}$ for temperature generated due to local shear~\cite{KimPRL2020}.

The colored symbols in Fig.~\ref{fig:fric_collapse} show local measurements of $\mu$ and $\Theta$ with $\Gamma = 3$, 10, 30, and 100 and $\tilde{A}$ varied over several orders of magnitude. As $\tilde{A}$ is increased (blue to pink), local measurements of $\Theta$ increase, spanning roughly five orders of magnitude, and local measurements of $\mu$ decrease. Following~\citet{KimPRL2020}, we also plot $\mu \Theta^{1/6}$ versus $I$ and obtain a collapse, excluding data with $\mu<0.1$ since the collapse breaks down as $\mu \rightarrow 0$.

We find similar results for $\mu_g = 0.1$ and $\mu_g = 0$, shown in Supplemental Material~\cite{suppmat}. When $\mu_g = 0$, we observe some variation in the exponents of the power-law scaling functions, as well as $\Theta \propto I$ for temperature generated purely due to local shear (no vibration), as opposed to $I^{3/2}$ as for frictional particles. 

Both the combinations of dimensionless variables, $\mu \Theta^{1/6} = F(I)$, and the functional form for $F(I)$ are very similar to those in Ref.~\cite{KimPRL2020}. This is a very surprising result, since in their case $\Theta$ was generated purely from shear and was primarily understood as fluctuating particle velocities during creeping flows generated by diffusing fluidity~\cite{Zhang2017PRL}. We generate $\Theta$ via external vibrations, which has no particular relationship to fluidity and thus isolates the role of temperature,  and span a much larger range of $\Theta$. Unexpectedly, the $\mu(I,\Theta)$ relation appears similar. We therefore conclude that friction can be predicted entirely based on the local values of $I$ and $\Theta$; no other information about the fields is necessary, including global measures like $\Gamma$ and $\tilde{A}$ or nonlocal measurements dependent on the values of $I$ or $\Theta$ in neighboring regions.

\textit{Modeling temperature---}As noted in Refs.~\cite{Zhang2017PRL,KimPRL2020}, this new field $\Theta$ must be modeled to close the equations. Despite the fact that diffusion of granular temperature is a standard part of hard-sphere kinetic theory \cite{campbell1997self, andreotti2013granular} and modifications to moderately dense flows exist \cite{jenkins2010dense,Berzi2024PRF}, there is currently no temperature-based theory that predicts both temperature and friction in dense, flowing granular matter that has persistent, elastic contacts. Existing equations for dense granular gases are inconsistent with our data; e.g., Eq.~(5) in~\citet{Berzi2024PRF} predicts $\Theta \propto I^2$ for no vibration, whereas we observe $I^{3/2}$ with grain-grain friction and $I$ with no friction. The discrepancy is likely due to the fundamental role of the sustained contact stresses in the dense system.

We propose a granular temperature diffusion equation with both a sink and source term, which captures many features of our data:
\begin{equation}
    \frac{\partial \Theta}{\partial t} = D \nabla^2 \Theta - B \Theta + A I^{a},
    \label{eqn:mod-heat-eq}
\end{equation}
where $D$ is a constant, diffusion coefficient, $B$ is a dissipation parameter arising from grain interactions, and $A$ and $a$ are a magnitude and exponent for a temperature source due to shear. 

The linear dependence on $\Theta$ in the sink term ($B\Theta$) is consistent with a viscous dissipation model where the energy is dissipated in proportion to the square of local velocity differences.  This is a distinct form from the kinetic theory for hard-sphere which has temperature-dependent conductivity and a sink term proportional to $T^{3/2}$ based on the collision rate and energy loss per collision \citep[e.g.,][p. 175]{Andreotti2013}. The source term $A I^a$  captures the temperature generation by shear. For frictional grains, $a = 3/2$ empirically (Fig.~\ref{fig:fric_collapse}), while for frictionless grains $a = 1$ (see Supplemental Material~\cite{suppmat}). Eq.~\eqref{eqn:mod-heat-eq} combined with appropriate boundary conditions can provide a closed system to describe the granular flow. 
 
As a first test of Eq.~\ref{eqn:mod-heat-eq}, we explore the temperature profiles of Figure~\ref{fig:cartoon}. For slow (very small $I$), steady flows with significant vibration, the source term for $I$ can be neglected and $d \Theta / dt = 0$. Symmetry in the $x$ and $y$ directions means $\nabla^2 \Theta \rightarrow \partial^2 \Theta / \partial z^2$, and the remaining equation is solved by a decaying exponential, $\Theta(z) \approx \Theta_0 \exp(-z/z_0)$, as observed in Fig.~\ref{fig:cartoon} for the limit of large vibration and slow shear. Here, $\Theta_0$ is set by the boundary condition at $z=0$ and $z_0 = \sqrt{D/B}$. We perform linear fits to $\log \Theta$ versus $z$ to obtain $\Theta_0$ and $z_0$. 

\begin{figure}
    \includegraphics[trim=0mm 0mm 0mm 0mm, clip,width=\columnwidth]{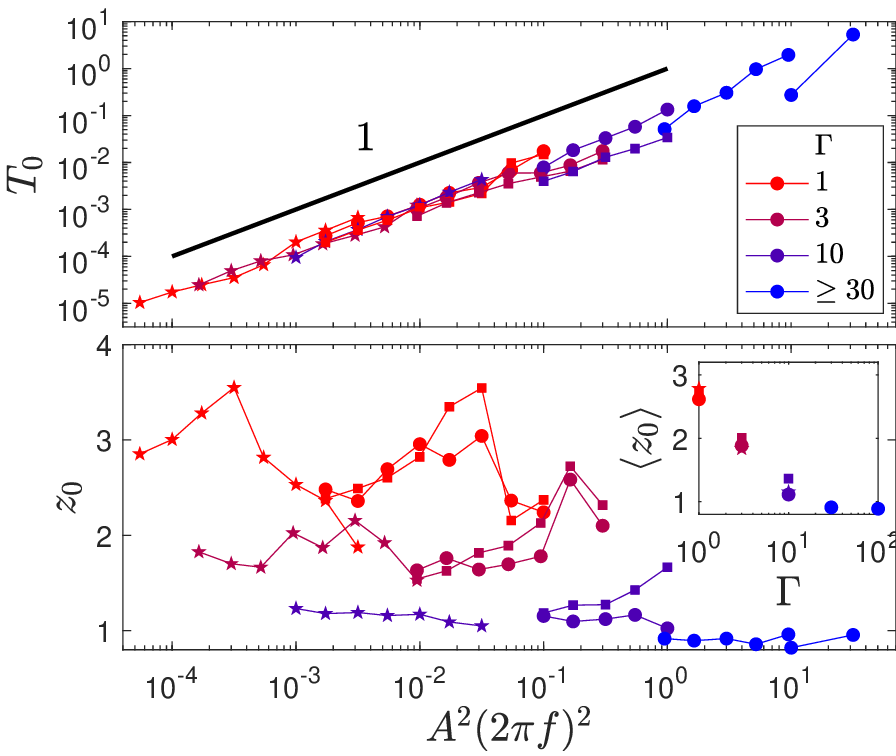}\\
    \caption{Dimensioned wall temperature $T_0 = \Theta_0 P/\rho$ and the decay length $z_0$ plotted as a function of the square of the vibration velocity, $A^2(2\pi f)^2$; the solid line shows $T_0 = A^2(2\pi f)^2$. Symbols represent $\tilde{p} = 10^{-4}$, $e_n = 0.5$ (circles); $\tilde{p} = 10^{-4}$, $e_n = 0.2$ (squares); and $\tilde{p} = 10^{-5}$, $e_n = 0.5$ (stars). The inset shows the average value of $z_0$ versus $\Gamma$.}
    \label{fig:decay_length}
\end{figure}

The boundary condition $\Theta_0$ is set by the square of the vibrating wall velocity.  Figure \ref{fig:decay_length}(a) shows a plot of the dimensioned wall temperature $T_0 \equiv \Theta_0 P/\rho$ versus $A^2(2\pi f)^2$, which is the velocity scale of the vibrating wall squared. The reference curve (black line) has $T_0 = A^2(2 \pi f)^2$, and the data are parallel to this curve.

The decay length $z_0$ is a constant, independent of $\tilde{A}$ and $\Gamma$ for $\Gamma\gg 1$. Fig~\ref{fig:decay_length}(b) shows that $z_0$ depends on $\Gamma$ only near the transition to contact breaking near $\Gamma = 1$, becoming constant for $\Gamma\geq 10$. Also note that the three different symbols in Fig.~\ref{fig:decay_length} represent varied $\tilde{p}$ and $e_n$. The fact that no meaningful difference is observed in $z_0$ is interesting and suggests that either $D$ and $B$ scale in the same way with $e_n$ or that grain-grain friction dominates loss. Future work is required to validate and better understand the terms in Eq.~\eqref{eqn:mod-heat-eq}.

\textit{Contact stress, fabric, and local rheology---}What grain-scale mechanisms connect local values of $\mu$, $I$, and $\Theta$? We first note that $\Theta$ can be thought of as a ratio of measures of kinetic stresses $\rho T$ and total stress $p$. Since the collapse in Fig.~\ref{fig:fric_collapse}(c) uses almost exclusively data with $\Theta < 10^{-2}$, we expect that $\mu$ is dominated by the contact stresses for our data.

This allows us to utilize previous work connecting $\mu$ to fabric anisotropy for contacts, $a_c$, for normal forces, $a_n$, and for tangential forces, $a_t$~\cite{bathurst1990observations,srivastava2020evolution}. These quantities range from 0 (perfectly isotropic) and 1 (perfectly anisotropic) and arise from a Fourier expansion of the distribution of contact angles and forces. For example, the probability distribution of contact vectors between particles can be approximated to first order as $\frac{1}{4\pi}(1+a_{ij}^c n_i n_j)$, where $n_i$ is the $i$-th Cartesian component of the branch vector. $a_c$ is the second invariant of $a_{ij}^c$, and similar first-order equations can be found for the distributions of normal and tangential forces to define $a_n$ and $a_t$. Previous work has shown that, to first order, $\mu$ from contact stresses can be approximated as $\mu = \frac{2}{5}\left(a_c + a_n + \frac{3}{2}a_t\right)$~\cite{bathurst1990observations,srivastava2020evolution}. 

\begin{figure}
    \raggedright (a)  \hspace{37mm} (b) \\
    \centering
    \includegraphics[trim=0mm 0mm 0mm 0mm, clip,width=0.49\columnwidth]{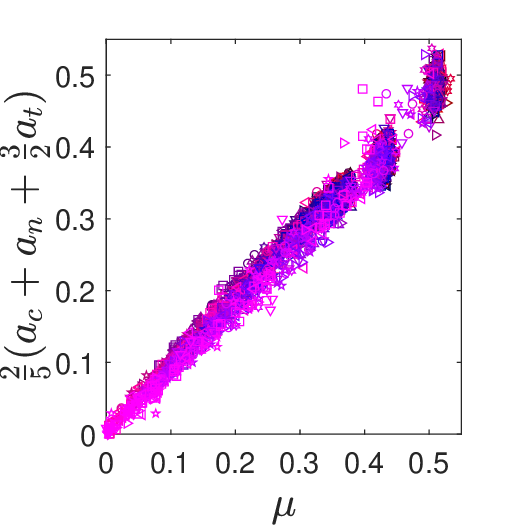}
    \includegraphics[trim=0mm 0mm 0mm 0mm, clip,width=0.49\columnwidth]{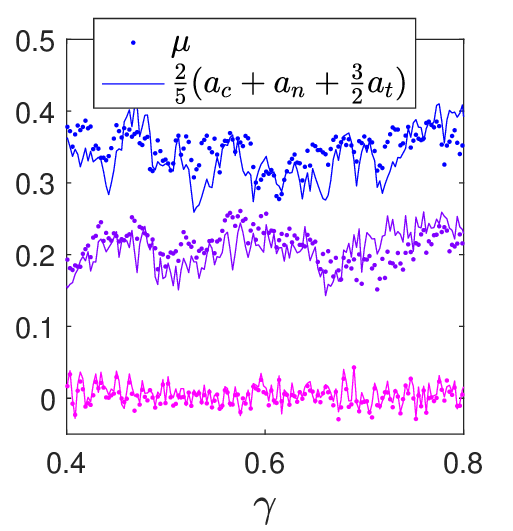}
    \centering
    \caption{(a) Local, time-averaged measurements of $\frac{2}{5}(a_c+a_n+\frac{3}{2}a_t)$~\cite{bathurst1990observations,srivastava2020evolution} plotted versus $\mu$ for data with $\Theta<0.1$. (b) Typical sequence during a simulation of $\mu$ and $\frac{2}{5}(a_c+a_n+\frac{3}{2}a_t)$ as a function of strain weak (blue), moderate (purple), and strong (pink) vibration}
    \label{fig:mu_proxy}
\end{figure}

We compute $a_c$, $a_n$, and $a_t$ using built-in functionality for LAMMPS. Figure~\ref{fig:mu_proxy} shows that $\mu \approx \frac{2}{5}\left(a_c + a_n + \frac{3}{2}a_t\right)$ even during frictional weakening associated with vibration, provided that $\Theta \ll 1$. We show time-averaged measurements for all data with $\Theta < 0.1$ in panel (a) and time series of both quantities for representative layers during a simulation in panel (b); both show good agreement. Thus, reduction in $\mu$ due to increased $\Theta$ is associated with reduced $a_c$, $a_n$, and $a_t$.

\begin{figure}
    \raggedright (a)  \hspace{37mm} (b) \\
    \centering
    \includegraphics[trim=0mm 0mm 5mm 0mm, clip,width=0.49\columnwidth]{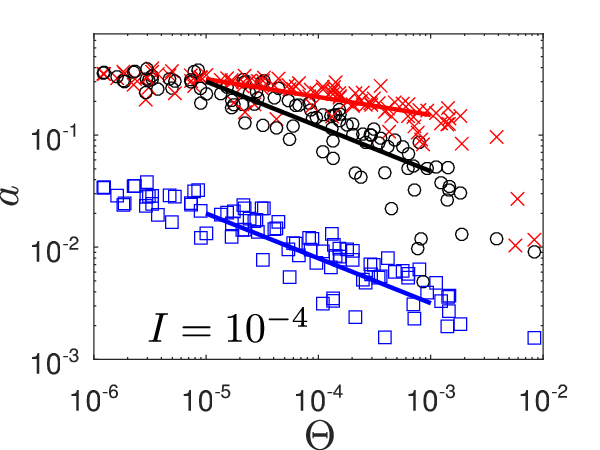}
\includegraphics[trim=0mm 0mm 5mm 0mm, clip,width=0.49\columnwidth]{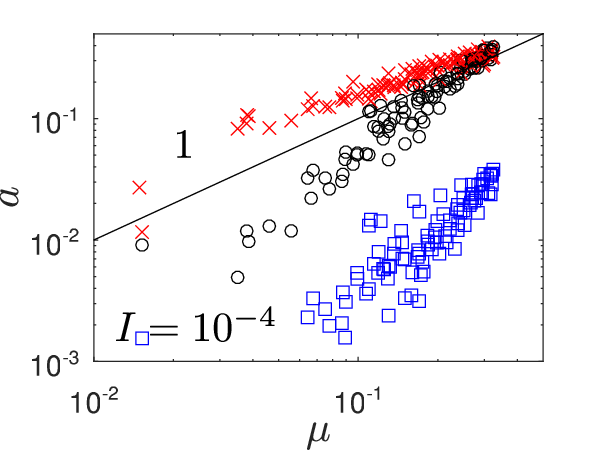}\\
    \raggedright (c)  \hspace{37mm} (d) \\
    \centering
    \includegraphics[trim=0mm 0mm 0mm 0mm, clip,width=0.49\columnwidth]{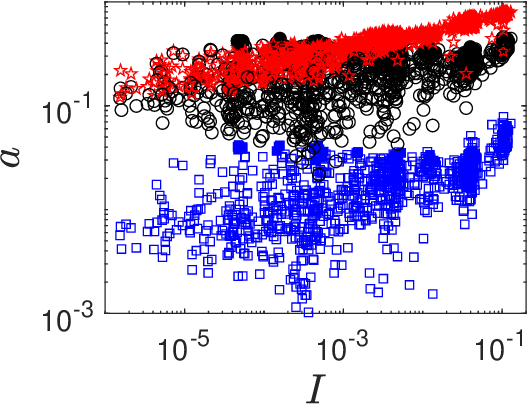}
    \includegraphics[trim=0mm 0mm 0mm 0mm, clip,width=0.49\columnwidth]{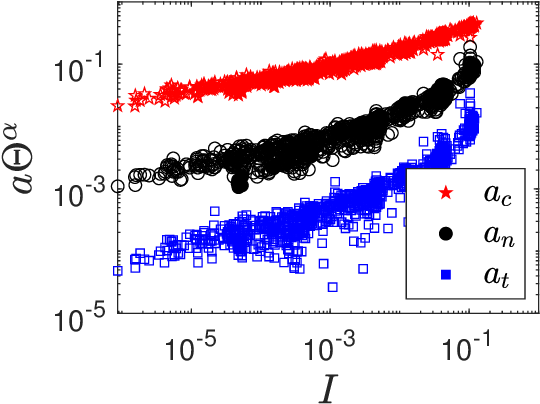}\\
    \centering
    \caption{(a) Local measurements of $a_c$ (red crosses), $a_n$ (black circles), and $a_t$ (blue squares) versus (a) $\Theta$ and (b) $\mu$, both with $I\approx 10^{-4}$ ($\pm 20\%$). Solid lines in (a) correspond to power laws with exponent $-1/6$ for $a_c$ and $-2/5$ for both $a_n$ and $a_t$. (c) Local measurements $a_c$, $a_n$, and $a_t$ (same symbol convention) versus $I$ for widely varied vibration conditions (all with $\Gamma>1$ and $\mu>0.1$ as in Fig.~\ref{fig:fric_collapse}). (d) $a_c \Theta^{1/6}$, $a_n \Theta^{2/5}$, and $a_t \Theta^{2/5}$ versus $I$, where the exponent values are taken from the solid lines in panel (a).}
    \label{fig:fabric_components}
\end{figure}

This raises the question of whether $\mu$ or the fabric components are more fundamental, i.e., whether a local rheology $a(I,\Theta)$ should instead be formulated for each of the fabric components. Figure~\ref{fig:fabric_components}(a) and (b) show $a_c$, $a_n$, and $a_t$ for local $I \approx 10^{-4}$; we find similar results for other values of $I$. As $\Theta$ is increased, all three decrease. The contact anisotropy $a_c$ falls of the slowest, while the force anisotropies $a_n$ and $a_t$ fall off faster. Physically, this means preferential directions for contacts are more easily maintained than for forces. This is reinforced by Figure~\ref{fig:fabric_components}(b), which has the same data as (a) but with $\mu$ on the horizontal axis instead of $\Theta$. As $\mu$ is decreased due to $\Theta$, $a_c$ decreases less rapidly than $a_n$ and $a_t$

 The solid lines in Fig.~\ref{fig:fabric_components}(a) show $a_c \propto \Theta^{-1/6}$, $a_n \propto \Theta^{-2/5}$, and $a_t \propto \Theta^{-2/5}$. These power-law functions appear to be approximately valid over a range of $\Theta$, but fail for very large $\Theta$, where especially $a_c$ drops precipitously, as well as very small $\Theta$, where the curves flatten out somewhat. Figure~\ref{fig:fabric_components}(c) shows $a_c$, $a_n$, and $a_t$ plotted versus $I$ for all values of $\Theta$, and panel (d) shows $a\Theta^\alpha$ plotted as a function of $I$, where $\alpha$ is $1/6$ for $a_c$, $2/5$ for $a_n$, and $2/5$ for $a_t$. Thus, the collapse of $\mu \Theta^{1/6} = F(I)$ is at minimum closely correlated to a similar collapse in these measures of fabric anisotropy.

\textit{Conclusions---}In summary, we have shown the existence of a local rheology for sheared, vibrated granular flows. Local granular temperature and local stresses are enough to predict time-averaged flow. Temperature profiles are consistent with a modified heat equation.   The local rheology we observe is consistent with that observed previously \citep{KimPRL2020}.  Additionally, previously studied relations between stresses and fabric tensors are still valid for the majority of our data, implying that a theory based on the contact and force fabric tensors may be appropriate. This work suggests that a temperature-based approach can provide a unified, local rheology of dense granular flow. 

\begin{acknowledgments}
We gratefully acknowledge funding from Army Research Office under grants W911NF1510012 and W911NF2220044 and the Office of Naval Research under grant  N0001419WX01519. We also thank Sachith Dunatunga for many helpful discussions in the early stages of this project, as well as Jeffrey Haferman and Michael Hernandez for help with high-performance computing at NPS.
\end{acknowledgments}

\bibliography{references}

\end{document}